\documentclass[structabstract]{aa}  
\usepackage{graphicx}
\usepackage{amsmath}
\usepackage{booktabs}
\usepackage{balance}
\usepackage{txfonts}
\usepackage{natbib}
\bibpunct{(}{)}{;}{a}{}{,} 
\usepackage{hyperref}
\setcitestyle{notesep={ }}
\begin{document}
   \title{Nuclear de-excitation line spectrum of Cassiopeia~A}

   \subtitle{}

   \author{A. Summa \and D. Els\"asser \and K. Mannheim}

   \institute{Institut f\"ur Theoretische Physik und Astrophysik, Universit\"at
							W\"urzburg, Campus Hubland Nord, Emil-Fischer-Str. 31, D-97074 W\"urzburg, Germany\\
              \email{asumma@astro.uni-wuerzburg.de}
             }

   \date{Received / Accepted}

 
  \abstract
   {The supernova remnant Cassiopeia A is a prime candidate for accelerating
cosmic ray protons and ions. Gamma rays have been observed at GeV and TeV
energies, which indicates hadronic interactions, but they could also be caused by
inverse-Compton scattering of low-energy photons by accelerated electrons.}
   {We seek to predict the flux of nuclear de-excitation lines from Cas A through
lower-energy cosmic rays and to compare it with COMPTEL measurements.}
   {Assuming a hadronic origin of the high-energy emission, we extrapolate the
cosmic ray spectrum down to energies of
$10\,\mathrm{MeV}$, taking into account an equilibrium power-law momentum
spectrum with a constant slope. We then calculate the nuclear line spectrum of
Cassiopeia A, considering the most prominent chemical elements in the MeV band
and their abundances as determined by X-ray spectroscopy.}
   {We show that the predicted line spectrum is close to the level of the
COMPTEL sensitivity and agrees with conservative upper limits.}
   {}


   \maketitle
%

\section{Introduction}

Cassiopeia A (Cas A) is one of the youngest known Galactic supernova remnants (SNR). 
Extensive observations in the radio, infrared, optical, and X-ray bands
lead to an estimated explosion date of around 1680 AD \citep{Thorstensen2001,
Fesen2006}. Being the brightest radio source in our Galaxy, it was first
discovered in 1948 \citep{Ryle1948}. Its distance can be estimated to
$3.4_{-0.1}^{+0.3}\, \mathrm{kpc}$ based on the combination of Doppler shifts
and proper motions, whereas the angular size of $2.5'$ corresponds to a physical
size of $2.34\, \mathrm{pc}$ \citep{Baars1977}. The optical spectrum obtained
from distant light echos of the original blast indicates that Cas A was a type
IIb supernova of a $\sim 15\,M_\odot$ main-sequence star and originated from the
collapse of the helium core of a red supergiant that had lost most of its
hydrogen envelope before exploding \citep{Krause2008}.
   
   The blast wave drives an initially outgoing forward shock that can be seen
as a thin X-ray edge expanding at roughly $5000\, \mathrm{km\, s^{-1}}$. The
interaction with circumstellar material and the interstellar medium results in a
reverse shock that is driven back into the outgoing ejecta and expanding at
roughly half the rate of the forward shock \citep{Gotthelf2001, DeLaney2003}.
Emission at most wavelengths is dominated by a $30''$ thick ``bright ring''
where the ejecta are heated and ionized when they encounter Cas A's reverse
shock. Viewed in X-rays, this shell consists of undiluted ejecta rich in O, Si,
S, and Fe \citep{Willingale2002, Laming2003, Lazendic2006}. Faint X-ray filaments
outside of the shell mark the location of the forward shock where nonthermal
X-ray synchroton radiation is produced by shock-accelerated electrons
\citep{Gotthelf2001, Vink2003}. Detected by HEGRA \citep{Aharonian2001}, MAGIC
\citep{Albert2007} and VERITAS \citep{Humensky2008}, Cas A was the first SNR
verified in TeV gamma rays. Recent observations with Fermi-LAT in the GeV range
do not rule out either a leptonic or a hadronic emission scenario: A combination
of nonthermal bremsstrahlung and inverse-Compton emission on the one hand as
well as neutral pion decays on the other hand could be responsible for the
measured spectrum \citep{Abdo2010}.
   
   It has long been suggested that the shock waves associated with supernovae
could be sites of acceleration of cosmic-ray particles \citep{Baade1934}. The
combination of diffusive shock acceleration (DSA) processes in SNR shocks with
transport effects in our galaxy can in theory produce the observed power-law
spectrum of cosmic rays up to the ``knee'' of about $10^{15}\, \mathrm{eV}$
\citep{Krymskii1977, Axford1977, Bell1978, Bell1978a, Blandford1978, Drury1983,
Jones1991}. To be consistent with the energy density of Galactic cosmic rays,
the transfer of kinetic energy released in supernova explosions to cosmic rays
must take place with an efficiency of $\sim\!10\,\%$ \citep{Ginzburg1969}.
According to the DSA model, the charged particles scatter off irregularities in
the magnetic field and increase their momentum by a fraction of the shock
velocity $v/c$ each round trip from the downstream to the upstream region and
back again. Because of the feedback of the accelerated particles on the spatial
profile of the flow velocity and therefore on the particle distribution itself,
the DSA process can be significantly nonlinear. High-energy particles with considerable
diffusion lengths sample a broader portion of the flow velocity profile and
therefore experience a greater change in compression ratio than low-energy
particles. That is why the particle spectrum and the resulting photon spectrum
tend to flatten with increasing energy \citep{Baring1999, Ellison2000, Bell2001,
Ellison2005}. The correspondence of electron and proton spectra at high energies
\citep{Ellison2000} renders it extremely difficult to prove the acceleration of
cosmic ray protons at SNR shocks directly. Though the presence of cosmic ray
protons could be inferred by gamma rays resulting from collisions with ambient
gas and subsequent pion decays, TeV gamma rays can also be produced by
inverse-Compton scattering of cosmic microwave background photons with the
accelerated cosmic-ray electrons in SNRs.
   
   The detection of nuclear de-excitation gamma-ray lines could point a way out
of this dilemma: The line spectra produced in energetic collisions of cosmic
rays with the ambient medium provide an unambiguous proof for the hadronic
scenario of cosmic ray acceleration in SNRs. In contrast to other proposed
scenarios \citep[cf.][]{Bozhokin1997}, we expect the excitation processes to
happen near the proton acceleration site, so effects through Coulomb losses
regarding the proton spectrum can be neglected here. The following sections are
devoted to a short summary concerning the theoretical concepts of nuclear
de-excitation processes as well as the derivation of the resulting gamma-ray
fluxes in the MeV range with respect to current and future detection limits.
Finally, we discuss the results and their implications in the last section.

\section{Theoretical framework} \label{sec:2}

Inelastic scattering of energetic particles on heavier nuclei as well as
spallation reactions leaving the product nucleus in an excited state are
followed by the emission of gamma rays in the range from 1 to $20\,
\mathrm{MeV}$ through de-excitation processes. The spectral structure of these
gamma rays is determined both by the composition and the energy spectrum of the
energetic particles and by the respective properties of the ambient medium.
Especially cosmic rays with energies less than $100\, \mathrm{MeV}$ are suited
well to be studied in gamma rays, because for cosmic rays with greater energies
the gamma fluxes owing to \textit{p-p} and \textit{p-$\alpha$} collisions
followed by $\pi^0$ desintegration are expected to be higher than the gamma
fluxes resulting from nuclear de-excitations \citep{Meneguzzi1975}. That is why
the observation of gamma-ray lines below $100\, \mathrm{MeV}$ offers the
opportunity of studying astrophysical processes in great detail and of revealing the
origin of hadronic cosmic rays in SNRs. 
   
   The basic ingredients in determining the profile of a gamma-ray line from
energetic particle interactions can be summarized as follows: According to
\citet{Ramaty1979}, the probability of photon emission per second into solid
angle $d(\cos\theta_0)\:d\phi_0$ can be written as (the $z$-axis is chosen in
the direction of the incident particle)
 	\begin{equation}
\begin{split}dP_{\gamma}=\ &
n_iv\frac{d\sigma}{d\Omega^*}(E,\theta_{r}^*)\:d(\cos\theta_{r}^*)\:d\phi_{r}
\:\times\\[3mm]
 & \times
g(E,\theta_{r}^*,\theta_{0},\phi_{r}-\phi_{0})\:d(\cos\theta_{0})\:d\phi_{0}
.\end{split}
 	\label{eq:1}
	\end{equation}	
In the center-of-mass frame, the interaction produces an excited nucleus with
recoil velocity in $d(\cos\theta_r^*)\:d\phi_r$. $E$ represents the energy of
the incident particle before the interaction, $n_i$ the number density of the
target particles. $v$ is the incident particle's velocity, $d\sigma/d\Omega^*$
is the center-of-mass differential cross section, and $g$ is  the angular
distribution of the resulting gamma rays. $\phi_r$ and $\phi_0$ are azimuth
angles measured in the $(x,y)$-plane, $\theta_r^*$ and $\theta_0$ are polar
angles given with respect to the z-axis. To finally determine the gamma-ray
spectrum, Eq. (\ref{eq:1}) can be integrated over $\cos\theta_r^*$, $\phi_r$,
$\cos\theta_0$ and $E$ by using the Monte Carlo simulation technique. Choosing
random numbers (uniformly distributed from 0 to 1), the integrations can be
carried out by solving for $\cos\theta_r^*$, $\phi_r$, $\cos\theta_0$ and $E$
from the equations
	\begin{align}
			R_{1}=\ & 
\frac{1}{\sigma(E)}\intop_{-1}^{\cos\theta_{r}^{*}}\frac{d\sigma}{d\Omega^{*}}
d\Omega^{*},\\
			R_{2}=\ & \frac{\phi_{r}}{2\pi},\\
			R_{3}=\ & C\intop_{0}^{E}vN_{p}(E')\sigma(E')dE',\\
			R_{4}=\ & \frac{1}{2}(1+\cos\theta_{0}).
	\end{align}
Here $N_p(E)$ is the number of incident particles per unit energy, $C$ is a
normalization constant. The probability of observing gamma rays of energies
between $E_\gamma$ and $E_\gamma+\Delta E_\gamma$ is then proportional to the
sum of all angular distributions $g$ for which $E_\gamma$ is in range. For a
detailed description of the outlined methods above and a deeper insight into the
different reaction types as well as the derivation of line production cross
sections, we refer the reader to \citet{Ramaty1979} and \citet{Kozlovsky2002}
and references therein.

\begin{figure*}
   \includegraphics[width=17cm]{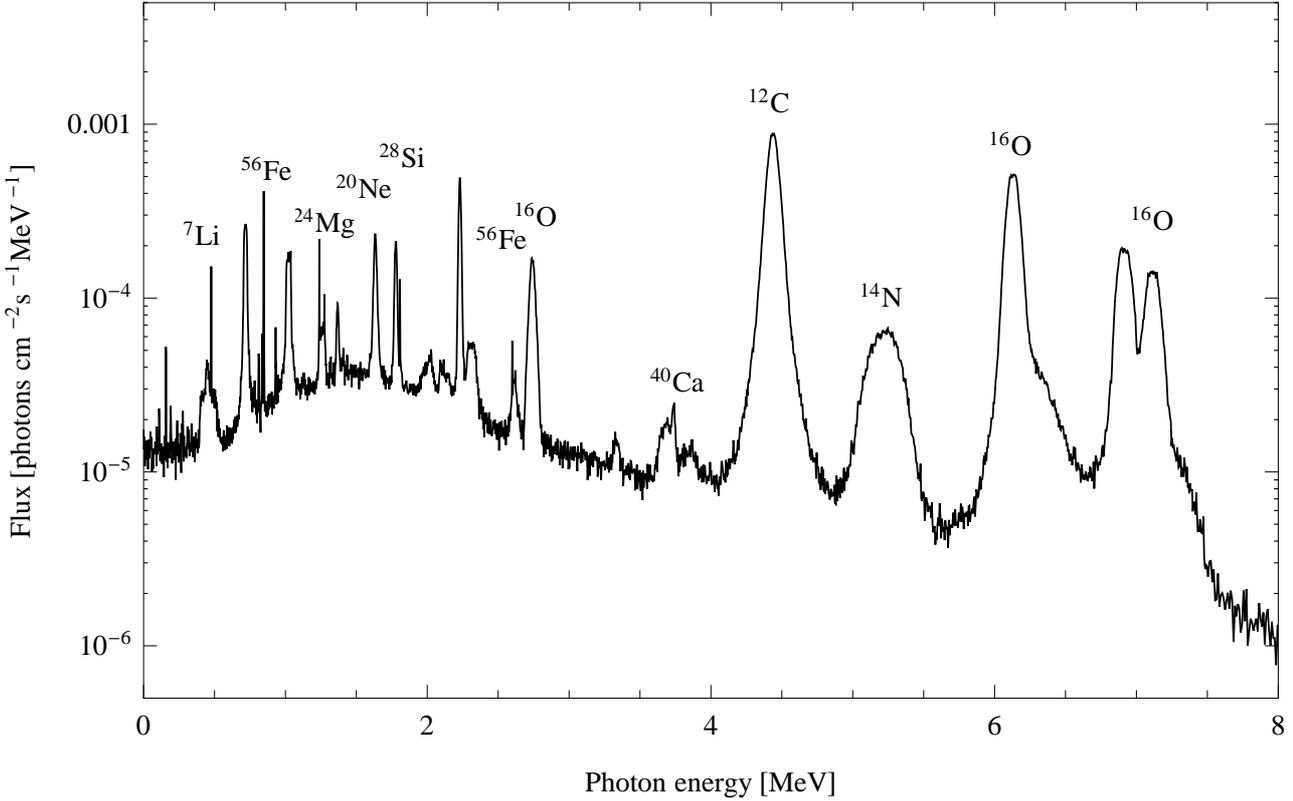}
   \caption{Calculated gamma-ray spectrum for the specific case of Cas A using
the assumptions described in the text. $10^6$ photons are binned into energy
intervals of widths ranging from 2 to 5 keV as described in \citet{Ramaty1979}.
For example, the $^{12}\mathrm{C}$ line at $4.4\, \mathrm{MeV}$ has a flux that
is comparable to the sensitivity of COMPTEL. Note that the continuum component
caused by nonthermal electron bremsstrahlung is not taken into consideration here.}
   \label{fig:1}
   \end{figure*}


\section{Results}

To validate the chances for success of the introduced approach concerning Cas A,
we first study the exemplary case of the carbon line at $4.4\,
\mathrm{MeV}$. Considering a scenario in which the gamma-ray emission of Cas A
is modeled by a hadronic fit based on the $\pi^0$-decay of accelerated hadrons,
the best-fit proton acceleration spectrum is given by $Q_p(p)\propto p^{-2.3}$.
The resulting proton energy content of $W_p=\int_{10\,
\mathrm{MeV}/c}Q_pp\,dp=4~\times~10^{49}\, \mathrm{erg}$ corresponds to $\sim2\,
\%$ of the estimated SNR kinetic energy \citep{Abdo2010}. By extrapolating the
high-energy proton spectrum down to the MeV-range, the gamma-ray flux emitted in
the $4.4\, \mathrm{MeV}$ nuclear de-excitation line can be approximately
calculated from
	\begin{equation}
	 F_{\gamma}=\frac{1}{4\pi d^{2}}n_{C}\intop Q_{P}(p)\sigma(p)v(p)dp,
	\end{equation}
where $n_C\sim 10\, \mathrm{cm^{-3}}$ is the adopted mean density of carbon
atoms in the interaction region \citep[cf.][]{Laming2003}, $v$ the velocity of
the accelerated protons and $\sigma$ the cross section for the inelastic
scattering processes. At first glance, line-broadening effects or additional
contributions from unresolved gamma-ray lines in heavy nuclei and lines from
long-term radioactive nuclei are neglected here. Using $d=3.4\, \mathrm{kpc}$
(see above) and the cross section for the reaction
$^{12}\mathrm{C}(p,p')^{12}\mathrm{C}^*$ given by \citet{Ramaty1979} yields a
flux of $\sim 10^{-6}\, \mathrm{cm^{-2}s^{-1}}$ at $4.4\,\mathrm{MeV}$. This
flux value is close to the sensitivity limit of the COMPTEL experiment: Following
the analysis of \citet{Strong2000} concerning the MeV continuum emission from
Cas~A, only an upper limit of $1.4\times 10^{-5}\, \mathrm{cm^{-2}s^{-1}}$ was
obtained in the $3-10\, \mathrm{MeV}$ energy range. According to
\citet{Iyudin1995}, the COMPTEL line sensitivity is indicated with $\sim
10^{-5}\,\mathrm{cm^{-2}s^{-1}}$. So only a future gamma-ray mission with
enhanced sensitivity in the MeV range will be able to obtain final results
concerning the detection of de-excitation lines in Cas~A.

   To compute the whole nuclear de-excitation spectrum for the specific case of
Cas A, we used the Monte-Carlo code developed by
\citet{Ramaty1979}\footnote[1]{\url{
http://lheawww.gsfc.nasa.gov/users/ramaty/code.htm}} (see also Sect.
\ref{sec:2}). Besides the ingredients already mentioned in the previous
paragraph, the calculations take into account the following assumptions: The
acceleration scenario for cosmic rays is assigned to the reverse-shock side 
and the chemical composition of the accelerated cosmic rays is inferred from 
\citet{Engelmann1990}. The composition of the ambient gas, in fact a mixture of both 
massive Wolf-Rayet winds and subsequent supernova ejecta \citep[cf.][]{Lingenfelter2007}, 
is described by the use of results 
from X-ray spectroscopy (cf. Table \ref{tab:1}). The abundances of H and He are 
deduced from optical measurements by \citet{Chevalier1979}. The resulting total mass is in 
line with the Wolf-Rayet-supernova scenario, i.e. there is no room for additional amounts of 
hydrogen that would enhance the pion vs. the nuclear de-excitation yields.
Unresolved gamma-rays from heavy nuclei and lines from long-term radioactive
nuclei are also included. Together with the consideration of recoiling target
particles, this leads to a significant broadening of the lines. Although some
nonthermal emission is associated with the forward shock, recent studies showed
that electron acceleration to multi-TeV energies is likely to take place mainly
at the reverse shock within the supernova ejecta \citep{Helder2008}, making Cas
A to a unique object for studying particle acceleration at the reverse-shock
side. The resulting spectrum is depicted in Fig. \ref{fig:1} and agrees well with the 
approximate calculation above.
   
\begin{table}
\caption{Mean measured abundance mass ratios and rms scatter resp. upper limits according to the
results of \citet{Willingale2002}, \citet{Docenko2010} and \citet{Chevalier1979}.}  
\label{tab:1}      
\centering                                      
\begin{tabular}{c c c}          
\toprule                        
ratio & mean & rms \\    
\midrule									
		H/Si & $<2.29\times 10^{-5}$ & -\\				
    He/Si & $<4.93\times 10^{-3}$ & -\\                               
    C/Si & 1.76 & 0.88 \\	
    O/Si & 1.69 & 1.37 \\
    Ne/Si & 0.24 & 0.37 \\
    Mg/Si & 0.16 & 0.15 \\
    S/Si & 1.25 & 0.24 \\
    Ar/Si & 1.38 & 0.48 \\
    Ca/Si & 1.46 & 0.68 \\
    FeL/Si & 0.19 & 0.65 \\
    FeK/Si & 0.60 & 0.51 \\
    Ni/Si & 1.67 & 5.52 \\
\bottomrule                                            
\end{tabular}
\tablefoot{The mass ratios are given relative to solar values. To faciliate the comparison, 
the data from \citet{Chevalier1979} for H and He are indicated relative to Si, too.}
\end{table}


\section{Discussion}

As can be seen in Fig. \ref{fig:1}, the adopted acceleration scenario for cosmic
rays in Cas A leads to a flux of nuclear de-excitation lines that would be
clearly detectable by a gamma-ray telescope with enhanced sensitivity as
sucessor to the COMPTEL mission \citep[e.g. the proposed GRIPS mission
by][]{Greiner2009}. The line emissivity is additionally boosted because the
thermal target gas reflects the heavily enriched abundances of the Wolf-Rayet
progenitor star. Though the detailed line characteristics always depend
on the precise knowledge of the supernova ejecta's composition, the natural
process of element synthesis in the progenitor star and the supernova explosion
mechanisms lead to peculiar properties in the gamma-ray spectrum. A unique
feature arises because the C and O lines in the $4-6\, \mathrm{MeV}$ band dominate
the line flux from the Ne-Fe group in the $1-3\, \mathrm{MeV}$ band. Of course,
the reasonability of extrapolating the proton spectrum obtained from high-energy
measurements down 
to the MeV range has to be discussed: It is commonly believed that strong
Coulomb losses would quench the particle spectra of nonthermal particles below
GeV energies, but the steep spectra of nonthermal particles in solar flares may
tell a different story. Furthermore, the results of modeling the acceleration
process of cosmic rays and the evolution of SNRs should be taken into account:
According to the kinetic models of \citet{Berezhko1997} and
\citet{Berezhko2000}, the overall proton momentum spectrum at the shock front is
close to a pure power law in the entire momentum range down to $\mathrm{MeV}$ at
low injection rates, for high injection rates the low-energy part $(p\ll mc)$ of
the overall cosmic ray spectrum is slightly steeper than the high-energy part.
The application of the kinetic model approach with respect to the specific case
of Cas A confirms these results \citep{Berezhko2003}. So the assumption of an in-situ 
cosmic ray spectrum that is as steep as at higher energies and a
localization of the inelastic scattering processes near the reverse shock site
seem to be quite reasonable and theoretically motivated.


\section{Conclusion}

We have shown that if cosmic rays are accelerated at the reverse shock in the
Wolf-Rayet supernova remnant Cas A, as indicated by its strong GeV and TeV
emission, the resulting flux of carbon and oxygen de-excitation lines is
marginally detectable with COMPTEL. Precision spectroscopy with future
MeV-missions such as GRIPS \citep{Greiner2009} will put the theory of cosmic ray
acceleration to a crucial test and will permit us to determine the overall
efficiency of cosmic ray acceleration.  As a corollary,
these measurements would also determine the yields of spallation products from
nuclear collisions that affect the abundances of the cosmologically relevant
light elements such as Li, Be, and B.  It must be noted that the prediction of
the line flux rests on an extrapolation of the cosmic ray spectrum determined at
high energies, which is in line with diffusive shock acceleration theory, but
which is still subject to major systematic uncertainties associated to nonlinear
effects such as shock broadening caused by back reactions.

\begin{acknowledgements}
The authors thank Werner Collmar for helpful discussions and the referee 
for the constructive comments on the manuscript. 
This work was funded by DFG through GRK 1147.
\end{acknowledgements}

\bibliographystyle{aa}
\balance
\bibliography{cas_a}

\end{document}